\documentclass{aa}
\usepackage{graphicx}

\newcommand {\GHz} {\ensuremath{ \mathrm{GHz} }}
\newcommand {\LogP} {\ensuremath{\log{P_{5\,\GHz} } } }

\begin{document}
\title{Milli-arcsecond scale Rotation Measure in the CSS
       Quasars 0548$+$165 and 1524$-$136}
\author{F.~Mantovani\inst{1} 
 \and W.~Junor\inst{2}
 \and R.~Ricci\inst{1,3} 
 \and D.J.~Saikia\inst{4} 
 \and C.~Salter\inst{5}
 \and M.~Bondi\inst{1}
}
\institute{Istituto di Radioastronomia, CNR, Via P. Gobetti 101, Bologna, 
Italy 
\and University of New Mexico, Albuquerque, New Mexico 87131, USA
\and Scuola Internazionale Superiore di Studi Avanzati, Via Beirut 4,
     34014 Trieste, Italy
\and National Centre for Radio Astrophysics, TIFR, Post Bag 3, Ganeshkhind, 
     Pune, 411007, India
\and Arecibo Observatory, HC3 Box 53995, Arecibo, Puerto Rico PR\,00612, USA
}
\offprints{Franco Mantovani,
\email{fmantovani@ira.bo.cnr.it}
}
\abstract{ 
Two Compact Steep-spectrum Sources (CSSs), 0548$+$165 and
1524$-$136, chosen from a list of CSSs with polarization
percentages that decrease with decreasing frequency
and high rotation measure values (RM$>450\,$rad\,m$^{-2}$) 
on arcsecond scales, were observed with the VLBA at 4.9 and
8.4\,GHz.  RM values up to $\approx10^4$\,rad\,m$^{-2}$ were found in
several regions along the jets in both sources. We suggest that a
thin screen of magneto-ionic material with about 1 kpc thickness is
responsible for these high RMs.  The observed depolarization
may be due to beam depolarization and/or inhomogeneities in the
magnetic field.
}
\date{Received / Accepted }
\titlerunning{mas scale RM in CSS Quasars }
\authorrunning{F. Mantovani et al. }
\maketitle

\section{Introduction}

Compact Steep-spectrum Sources (CSSs, e.g. Pearson {\it et al.} 1985; 
Fanti {\it et al.}
1990) are physically small objects with sub-galactic dimensions.
Their structures and sizes are possibly affected by the ambient gas in the 
central regions of the parent optical objects. At sub-arcsecond resolution 
these sources often show strongly distorted structures with
recognizable jet-like features (Fanti et al. 1986; Spencer 
et al. 1991), consistent with strong dynamical interactions between
the jets and the ambient media.  Most CSSs show low ($\sim$ 1$\%$)
percentage polarization at or below 5 GHz (Saikia 1991; Saikia, Singal
\& Cornwell 1987). However, the median polarization increases with
frequency (van Breugel et al. 1992; Saikia 1995, Saikia et al. 2001)
suggesting that the observations at lower frequencies are affected by
Faraday depolarization.  A number of CSSs have very high RMs, of the order
of several thousand rad\,m$^{-2}$.  To explain both the small linear
sizes and the low polarizations at centimeter wavelengths, it has
been postulated that CSSs are cocooned in dense gaseous envelopes ({\it
e.g.} Mantovani et al. 1994).  Supporting evidence for such envelopes
comes from optical spectroscopy (O'Dea 1998).

The largest structural distortions in CSSs are often seen in sources
dominated by jets, though this may be due in part to a selection
effect caused by projection effects (e.g. Spencer 1994).  However,
images of CSSs do indicate that intrinsic source distortions are
caused by interactions with a dense, inhomogeneous gaseous
environment.  This view is supported by the fact that the most
distorted and complex structures are found in objects with very weak
cores.  Additionally, apparent superluminal motions are rare among
CSSs, the only known cases to exhibit such motions being 3C147
(Alef et al. 1990) and 3C138 (Cotton et al. 1997).

In addition to distorted structures, increased asymmetry in the
location of the outer components also suggests that these components
are evolving through a dense asymmetric environment in the central
regions of the galaxies (Sanghera et al. 1995; Saikia 1995).
Sub-arcsecond polarimetry has provided evidence in favour of the
interaction of these components with dense clouds of gas. For example,
the southern component of 3C147 has a much higher RM than the
northern component, is brighter, closer to the nucleus, and has the
expected signature of a jet colliding with a cloud of gas on the
southern side of the galaxy (Junor et al. 1999).

As noted by Murgia et al. (1999), about 10-15\% of CSSs, mainly quasars, 
are jet-dominated with complex or highly asymmetric structures and may
represent a different class of ``frustrated'' objects where
strong jet-cloud interactions are at work.  

In most of the core-jet structures, the derived magnetic field
configurations are  parallel to the source major axis, as is found for
VLBI-scale jets in quasars (Cawthorne {\it et al.} 1993). In core-dominated
QSOs, high rest-frame Faraday rotation measures ($>10^3$ rad m$^{-2}$) are
found in the inner 20 pc of the nuclear region (Taylor 1999). 

Polarimetric observations of CSSs provide a useful probe of the
physical conditions in the gaseous environment of these young
sources. However, the number of CSSs for which detailed polarization
information is available is still rather small.  To improve this
situation, we are conducting a series of observations to image
several sources from a list of CSSs having reasonable degrees of
polarized emission and clear signatures of interaction between radio
emission and the environment, namely fractional polarizations that
decrease with increasing wavelength, and high values of rotation
measure (RM $>$450 rad m$^{-2}$ in the source rest frame).

\subsection{The two target sources}

So far, most of the sources observed in our candidate list are
QSOs.  The resulting images will be useful for a direct comparison
with the polarization structures found in core-dominated QSOs.   Here
we will present and discuss new VLBA polarization observations of the
two CSS quasars,  0548$+$165 and 1524$-$136.

In  arcsecond-scale images, 0548$+$165 shows an asymmetric structure
with a strong unresolved component coincident with a quasar at
redshift 0.474, and a faint secondary component 3 arcsec northward.
Most of the polarized emission comes from the main component. This
emission is strongly depolarized going from 15 to 5\,GHz, with a
rest-frame rotation measure of 1934 rad m$^{-2}$ (Mantovani et al.,
1994). The magnetic field direction derived from the derotated
electric vectors is aligned almost  east-west,
which is also the direction of the inner part of the
milli-arcsecond jet. Images obtained with the European VLBI Network
at 1.6 and 5\,GHz (Mantovani et al., 1998) show a collimated thin jet
with a noticeable bend at 80\,mas ($\simeq 253$\,pc) from the core.
However, the jet does not lose its collimation after the bend and can
be tracked for an additional 75\,mas. It is worth mentioning that
while the faint component is located 3 arcsec northward, the
jet points South.

1524$-$136 has a steep spectrum classified as C$^-$ ({\it i.e.}
inverted at low frequency) by Steppe et al. (1995) from observations
made in the range 90\,MHz--230\,GHz. The Einstein 0.2--4.5 keV  X-ray
luminosity is  2.7 $10^{46}$ erg\,sec$^{-1}$ (Wilkes et al. 1994).  Images 
from VLA observations of 1524$-$136 (Mantovani et al. 1994) show a rather
compact object which is resolved into a double source, at 15\,GHz.
The separation between the two components is $\sim$330\,mas. 
1524$-$136 has a rest-frame RM=$-$840\,rad\,m$^{-2}$  and does
not depolarise between 15 and 5\,GHz.

Table\,1 summarizes the physical parameters derived for the two sources
from these previous observations. The source name is followed by the
Optical Identification (OI), Redshift, Apparent Visual Magnitude,
log(Spectral Power) at 5\,GHz in W Hz$^{-1}$, Spectral Index ($S_{\nu}
\propto \nu ^{-\alpha}$), Linear Size (LS) in pc (H$_0$ =
100\,km\,s$^{-1}$ Mpc$^{-1}$ and q$_0$ = 0.5).  
%
%
\begin{table}
\begin{center} \caption{The observational parameters of the observed sources}
\begin{tabular}{ccccccc} \hline Source     & OI & $z$   & $m_v$ &
\LogP  & $\alpha$ & LS (pc)  \\ 
\hline 0548$+$165 & Q  & 0.474 & 17.0 & 26.46  &   0.5   &  330 \\
1524$-$136 & Q  & 1.687 & 21.0  & 27.75  &   0.7   &  475 \\ 
\hline
\end{tabular} 
\end{center} 
\end{table} 

\section{Observations and data analysis}
The observations were carried out with the VLBA 
\footnote {The Very Long Baseline Array and the Very Large Array are 
facilities of the National Radio Astronomy Observatory, USA, operated by 
Associated Universities Inc., for the National Science Foundation.} 
and one antenna of the VLA,
recording both right- and left-circular polarization (RCP, LCP) and using 
1-bit sampling. The observations are summarized in
Table\,2. Amplitude calibration was derived using both measurements
of the system temperatures made during the observations and
knowledge of the antenna gains of each element of the array.
Complex correlation coefficients were recovered at the Array
Operations Center, Socorro (NM, USA) correlator.  The recovered
complex correlation coefficients are corrupted principally by phase
gradients in frequency and time.  These are corrected by the use of
global fringe-fitting (Schwab \& Cotton, 1983).  Polarization
calibration was performed following Cotton (1993).  Calibration and
subsequent imaging were done in NRAO's ${\cal AIPS}$ analysis
package. The RCP$-$LCP delay difference was derived by fringe-fitting a
short segment of the cross-hand data from a strong calibrator.
Strong calibrators (DA193, OQ208, 3C84, 3C138, 3C286, 3C273, 3C345)
were used to determine the instrumental polarizations of the antennas. 
The polarization angle calibration used observations
of 3C286 at 8\,GHz and of 3C273 at the four frequencies in the 5-GHz
band.  Low resolution images of 3C286 were made from the calibrated
8.4-GHz data using the shortest baselines in the array. The
polarization position angle (PA) of the electric vector of the
linearly polarized emission was then compared with that measured in
VLA images of 3C286.  The 8.4-GHz data for each source, calibrated in
complex amplitude and phase, were edited and averaged in frequency and
time to improve the signal-to-noise ratio.  These data were
subsequently used in the standard iterative self-calibration and
imaging process.  Polarized images of un-polarized calibrator sources
were made to check our residual instrumental polarization.   These
images showed no systematic polarization miscalibration and the peak
polarization signal was $<$0.3\% of the peak total intensity signal.

Due to logistical problems, a slightly different PA calibration
strategy was used at 5\,GHz.  Here, four images, one for each of the
observing frequencies, were made of the source 3C273. This source has
a small rotation measure (RM$=80$ rad m$^{-2}$; Taylor et al. 1998)
for the strongest polarized component along the jet. We then derived
the PA values  for each frequency from a small box (5$\times$5 pixels)
around the peak of the polarized emission along the jet of 3C273.  The
LCP-RCP phase differences were corrected such as to place the PAs 
on a slope of 80\,rad\,m$^{-2}$ in a plot of PA against
$\lambda^{2}$.  The raw results from our analysis are shown in
Fig.\,1.  The required corrections applied to the PA for each IF were
as follows:  $\Delta$PA$_{4615}=6.9^\circ$,
$\Delta$PA$_{4653}=-45.5^\circ$, $\Delta$PA$_{4850}=-3.5^\circ$,
$\Delta$PA$_{5090}=0.8^\circ$.  The calibrated data were then imaged
and the polarization PA recovered.  These agree well with the nearly
contemporaneous integrated values in the UMRAO database \footnote{This
research has made use of data from the University of Michigan Radio
Astronomy Observatory which is supported by funds from the University
of Michigan.} (UMRAO 2001).
%
%
%
\begin{figure*}
\centering
\includegraphics[width=0.70\textwidth,bb=100 400 470 680,clip]{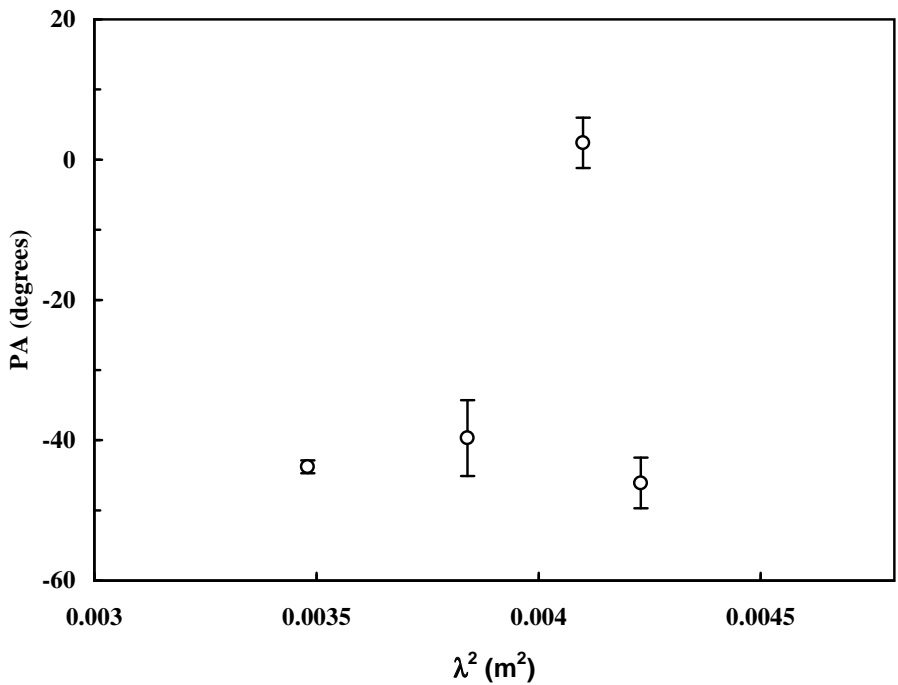}
\caption{PA {\it vs} $\lambda^2$ for the calibrator 3C273. }
\end{figure*} 
%
%
\begin{table*}
\begin{center}
\caption{A summary of the VLBA observations . Column 4 lists the observing
bandwidths, while Column 5 gives the raw integration times. The other columns
are self-explanatory. Note that the VLA1 was
available at only 4615 and 4850\,MHz. 
\label{vlbaobs} }
\vspace{0.5cm}
\begin{tabular}{llcccc}
\hline
Source     & Observing& Observing Frequency      & Bw   & $\Delta$ t \\
           &  Date    &      MHz                 & MHz  & sec        \\
\hline
0548$+$165 & 21JAN96  &  8405                    & 32   & 2  \\
DA193      &          &  8405                    & 32   & 2  \\
3C138      &          &  8405                    & 32   & 2  \\
\hline
1524$-$136 & 04FEB96  &  8405                    & 32   & 2  \\
OQ208      &          &  8405                    & 32   & 2  \\
3C286      &          &  8405                    & 32   & 2  \\
3C273      &          &  8405                    & 32   & 2  \\
\hline
0548$+$165 & 15FEB98  &  4615, 4653, 4850, 5090  & 8    & 4  \\  
1524$-$136 & 15FEB98  &  4615, 4653, 4850, 5090  & 8    & 4  \\  
OQ208      &          &  4615, 4653, 4850, 5090  & 8    & 4  \\
3C273      &          &  4615, 4653, 4850, 5090  & 8    & 4  \\
3C84       &          &  4615, 4653, 4850, 5090  & 8    & 4  \\
\hline
\end{tabular}
\vspace{0.5cm}
\end{center}
\end{table*}
\section{Results}
\subsection{Total intensity and polarization images}
Naturally-weighted images were made separately at the five observing
frequencies.  Natural weighting  gives the best sensitivity at the
expense of some resolution.  To allow direct comparison between the
images at different frequencies, the images were then convolved to a
common effective resolution via a two dimensional,
circularly-symmetric Gaussian restoring beam with a half-width of 6
mas.  The r.m.s. noises in the linearly polarized intensity
images are 0.15 and 0.4 mJy\,beam$^{-1}$ at 5 and 8.4\,GHz
respectively.  These values are close to the theoretically expected
noises.

As examples, Figs.\,2 and 3 show respectively the total intensity
and linearly polarized intensity images obtained at 4619\,MHz
for 0548$+$165, while Figs.\,4 and 5 display the images for
1524$-$136 at 4854\,MHz. For the latter source, we concentrate
here on the bright polarized jet-like component to the North. The
wide field images at 8.4\,GHz will be presented elsewhere.

The polarized and total flux density for five regions along the jet of 
0548+165 (see Fig.\,2) have been measured and are presented in Table\,3
together with  the S$_{pol}$ / S$_{tot}$ ratio for each
region. The Q map and the U map were integrated over a region separately
and then the polarized flux density, S$_{pol}= (Q^2+U^2)^{0.5}$, derived.
S$_{tot}$ and S$_{pol}$ are in mJy, the frequency $\nu$ in
GHz. In regions 2 and 3 the polarized flux density is below the 3-$\sigma$ 
noise level.
The same information for 1524$-$136 has been 
collected in Table\,4 for the regions marked A-F on Fig.\,5.
%
%
\begin{figure*}
\centering
\includegraphics[width=12cm]{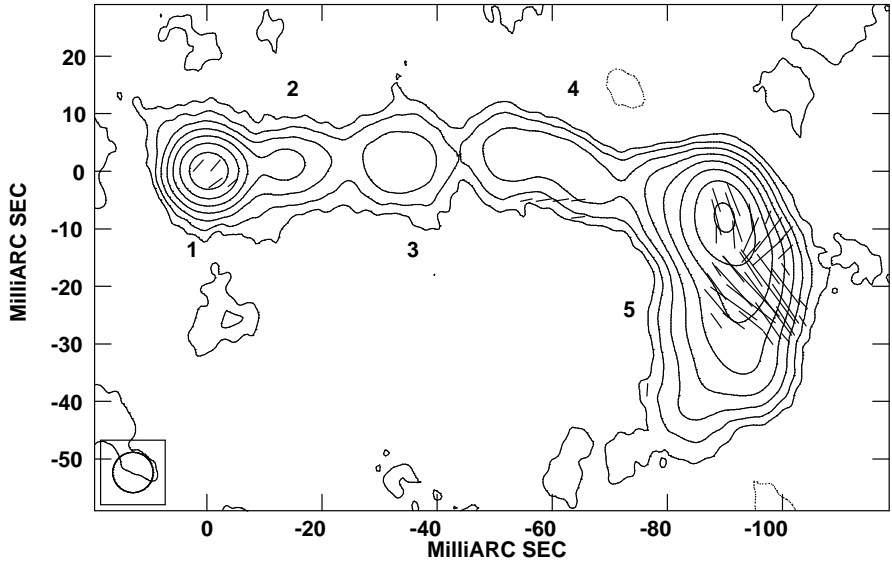}
\caption{The VLBA total intensity image of 0548+165 at 4.619 GHz. 
The contours are
at --1, 1, 2, 4, 8, 16, 32, 64, 128 mJy/beam. An electric field vector
length of 1 mas = 0.2 mJy/beam. The beam size is 6 $\times$ 6
mas and the peak brightness is 142.9 mJy/beam. The linear scale on the map 
is 1 mas = 3.5 $h_{0}^{-1}$ pc. \label{0548i}  
} 
\end{figure*}
%
%
\begin{figure*}
\centering
\includegraphics[width=12cm]{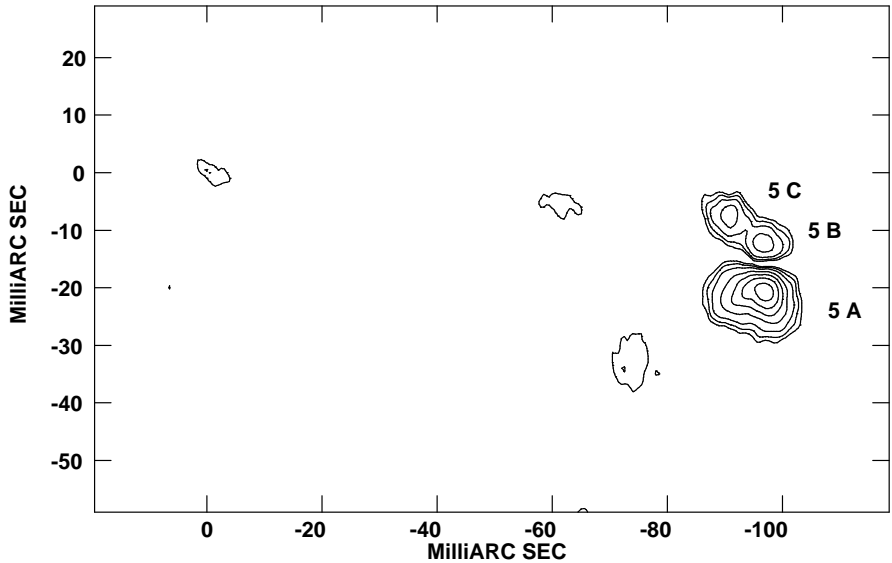}
\caption{The VLBA image of the linearly polarized intensity for 0548+165 
at 4.619 GHz. 
Contours are at --0.5, 0.5, 0.7, 1.0, 1.5, 2.0, 2.5, 3.0, 3.5 mJy/beam. 
The peak of the polarized emission is 3.3 mJy/beam. \label{0548pol}
} 
\end{figure*}
%
%
\begin{figure*}
\centering
\includegraphics[width=12cm]{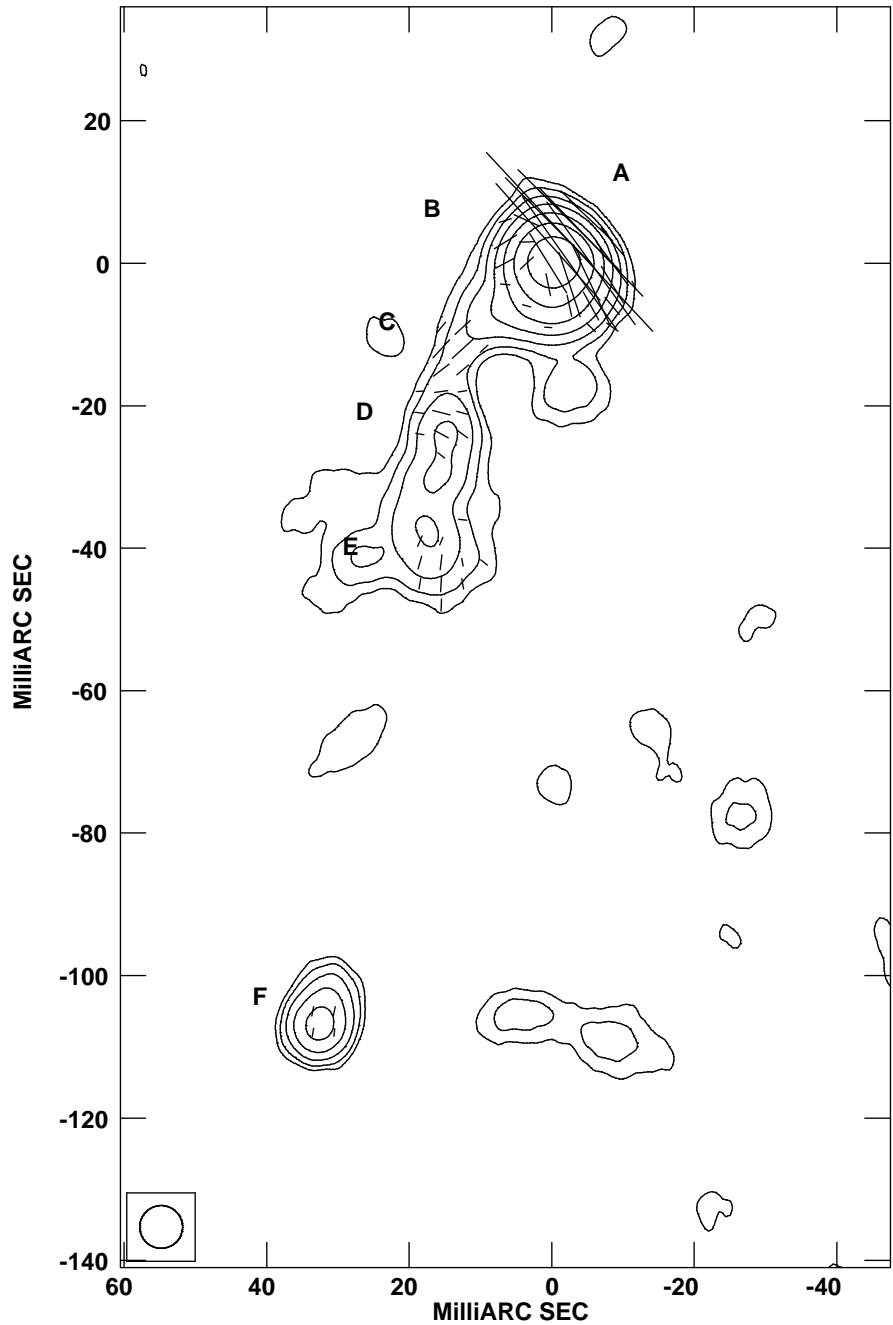}
\caption{The VLBA total intensity image of 1524$-$136 at 4.854 GHz. 
The contours are at --2, 2, 4, 8, 16, 32, 64, 128 mJy/beam.
An electric field vector length of 1 mas = 0.5 mJy/beam.
The beam size is 6 $\times$ 6 mas and the peak brightness is 217.9 mJy/beam.
The linear scale on the map is 1 mas = 4.2 $h_{0}^{-1}$
pc. 
         \label{1524i}
} 
\end{figure*}
%
%
\begin{figure*}
\centering
\includegraphics[width=12cm]{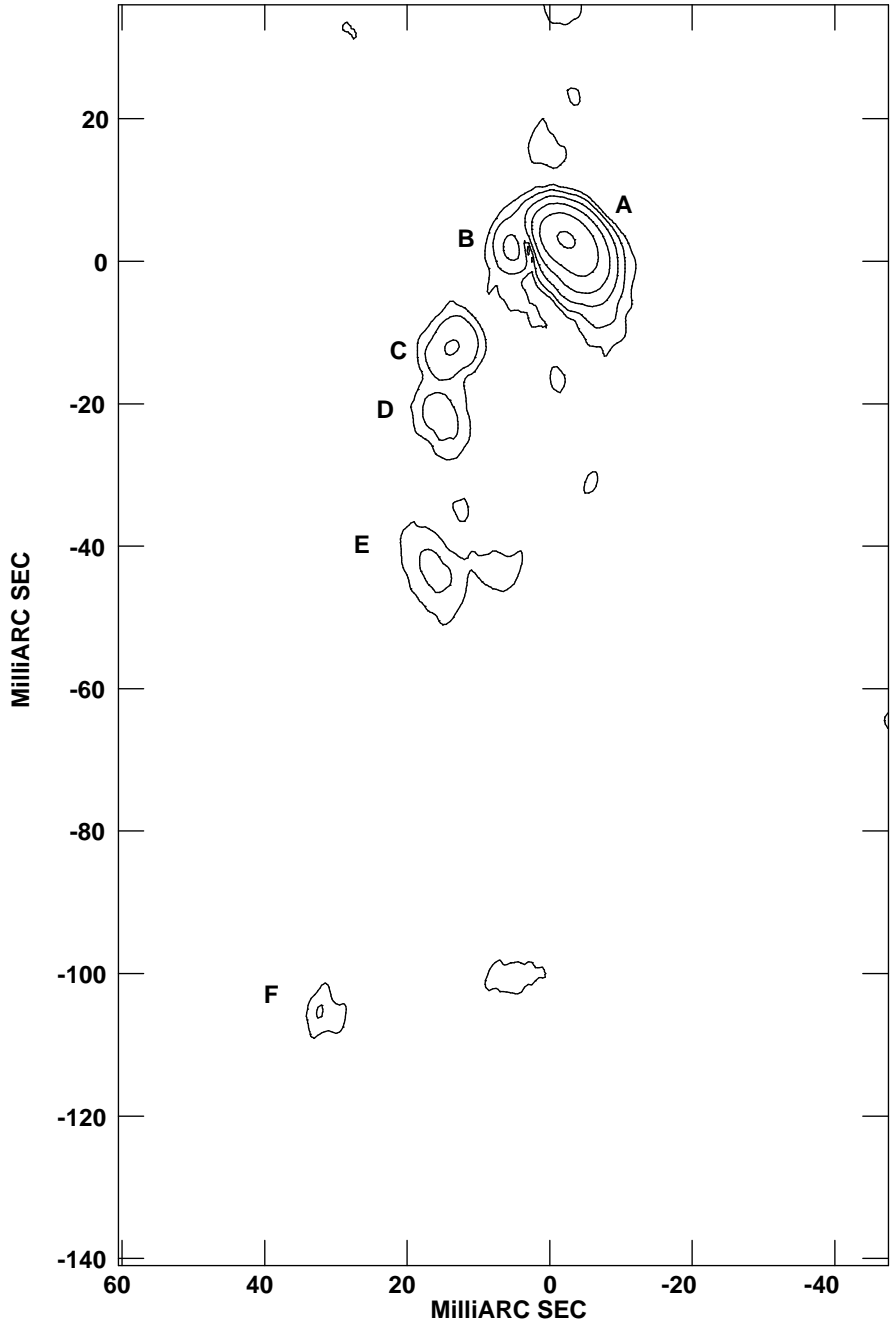}
\caption{The VLBA image of linearly polarized intensity for 1524$-$136 at 
4.854 GHz.
Contours are at --0.5, 0.5, 1, 2, 4, 8, 16 mJy/beam. 
The peak of the polarized emission is 17.1 mJy/beam.
\label{1524pol}  }  
\end{figure*}
%
%
\begin{table*}
\begin{center}
\caption{The total flux density and linearly polarized flux density for
five regions in 0548+165
         \label{0548info1} }
\begin{tabular}{cccccccccccccccc}
\hline
 & \multicolumn{3}{c}{Region 1} & \multicolumn{3}{c}{Region 2} &
    \multicolumn{3}{c}{Region 3} & \multicolumn{3}{c}{Region 4} & \multicolumn{3}{c}{Region 5} \\
\hline
 $\nu$ & S$_{tot}$ &  S$_{pol}$ & \% & S$_{tot}$ &  S$_{pol}$ & \% &
 S$_{tot}$ &  S$_{pol}$ & \% & S$_{tot}$ &  S$_{pol}$ & \% & S$_{tot}$
 &  S$_{pol}$ & \% \\ 
\hline
  4.615 & 115.4 & 0.6 & 0.5 & 9.6 & --- & --- & 7.4 & --- & --- & 12.2 & 0.6 & 4.9 & 501.1 & ~8.0 & 1.6 \\
  4.653 & 113.2 & 0.3 & 0.3 & 6.1 & --- & --- & 4.3 & --- & --- & ~8.6 & --- & --- & 365.3 & ~7.3 & 2.0  \\
  4.850 & 112.2 & --- & --- & 6.7 & --- & --- & 8.4 & --- & --- & 20.3 & 0.3 & 1.4 & 421.0 & ~8.8 & 2.1 \\
  5.090 & 112.8 & --- & --- & 4.5 & --- & --- & 5.7 & --- & --- & ~3.6 & 0.2 & 6.4 & 316.8 & 10.1 & 3.2   \\
  8.405 & ~86.0 & 0.5 & 0.6  & 1.8 & --- & --- & 2.0 & --- & --- & ~5.2 & --- & --- & 237.5 & 14.7 & 6.2  \\
\hline
\end{tabular}
\end{center}
\end{table*} 
%
%
\begin{table*}
\begin{center}
\caption{The total flux density and linearly polarized flux density for the 
five regions in 1524$-$136
\label{1524info1} }
\begin{tabular}{cccccccccccccccc}
\hline
 & \multicolumn{3}{c}{F} & \multicolumn{3}{c}{E} &
 \multicolumn{3}{c}{D} & 
    \multicolumn{3}{c}{C} & \multicolumn{3}{c}{B+A} \\
\hline
 $\nu$ & S$_{tot}$ &  S$_{pol}$ & \% & S$_{tot}$ &  S$_{pol}$ & \% &
 S$_{tot}$ &  S$_{pol}$ & \% & S$_{tot}$ &  S$_{pol}$ & \% & S$_{tot}$ &  S$_{pol}$ & \% \\
\hline
  4.615 & 54.1 & 1.5 & 2.8 & 75.7 & 3.8 & ~5.0 & 27.9 & 2.0 & ~7.1 & 17.9 & 3.5 & 19.6 & 539.6 & 28.1 & 5.2 \\
  4.653 & 46.7 & 1.0 & 2.1 & 31.2 & 1.6 & ~5.3 & 6.4 & 1.1 & 17.5 & 8.8 & 3.4 & 38.5 & 433.6 & 25.8 & 6.0  \\
  4.850 & 52.2 & 1.0 & 1.9 & 38.8 & 2.5 & ~6.5 & 18.7 & 1.5 & ~8.0 & 13.0 & 2.4 & 18.1 & 464.6 & 29.8 & 6.4 \\
  5.090 & 41.9 & 1.0 & 2.4 & 25.0 & 1.9 & ~7.7 & 11.2 & 0.9 & ~7.9 & 4.3 & 2.0 & 46.4 & 390.4 & 26.8 & 6.9   \\
  8.405 & 42.8 & 1.3 & 3.0 & ~8.6 & 1.0 & ~12.1 & ~6.5 & 4.1 & 63.5 & 6.8 & 2.1 & 30.2 & 200.1 & 16.2 & 8.1  \\
\hline
\end{tabular}
\end{center}
\end{table*}
\subsection{Rotation Measures}
Three regions of linearly polarized emission, labelled 5A, 5B and 5C on 
Fig.\,3, have been taken 
into consideration for the source 0548+165. The PA values for each of the IFs
have been measured considering the distribution of values in a small box
(5$\times$5 pixels) around the peak of the polarized emission. The standard
deviation for each value was also derived and added in quadrature to the 
error in the
PA correction applied by the calibration procedure 
(3$^\circ$). The PA values and their standard deviations from the mean are
given in Table\,5, together with the total flux densities, the linearly 
polarized flux
densities and the polarization percentages of the emission from the selected 
regions.

In Table\,6 the same information have been collected for  
six regions, A-F, within 1524$-$136. It should be noted that two of the 
four images made at the different C-band frequencies had
better uv coverages, and consequently a higher dynamic ranges, due to
the presence in the array 
of a VLA antenna. This causes the differences in the total flux densities
for each region of the two sources seen between the four C-band frequencies 
listed in Tables\,3 and 4.  
%
%
\begin{table*} 
\caption{Polarization Parameters for 0548+165 \label{0548info2}}
\vspace{0.5cm}
\begin{tabular}{lccccccc}
\hline 
$\nu$&$\lambda^2$& \multicolumn{2}{c}{Region 5A} &
\multicolumn{2}{c}{Region 5B} & \multicolumn{2}{c}{Region 5C} \\
\hline
MHz& m$^2 $& PA(deg)& $\sigma$ & PA(deg)& $\sigma$ & PA(deg)& $\sigma$ \\
\hline
4615& 0.00423 & 45.6 & 5.3 & $-$33.6 & 7.4 & ~22.7 & 6.1  \\
4653& 0.00410 & 47.1 & 5.3 & $-$10.6 & 10.0 & ~30.7 & 5.8  \\
4850& 0.00384 & 51.0 & 5.5 & $-$19.4 & 7.6 & ~71.0 & 5.3  \\
5090& 0.00348 & 60.6 & 6.2 &~~0.1 & 5.6 & 110.7 & 4.9    \\
8405& 0.00127 & 132.9 & 1.8 & ~99.4 & 1.8 & 355.2 & 1.4  \\
\hline
\end{tabular}
\end{table*}
%
%
\begin{table*} 
\caption{Polarization Parameters for 1524$-$136 \label{1524info2}}
\begin{tabular}{lccccccccccccc}
\hline 
$\nu$&$\lambda^2$& \multicolumn{2}{c}{Region A} &
\multicolumn{2}{c}{Region B} & \multicolumn{2}{c}{Region C}&
  \multicolumn{2}{c}{Region D} 
  & \multicolumn{2}{c}{Region E}& \multicolumn{2}{c}{Region F} \\
\hline
MHz& m$^2 $& PA(deg)& $\sigma$ & PA(deg)& $\sigma$ & PA(deg)& $\sigma$ 
& PA(deg)& $\sigma$ & PA(deg)& $\sigma$ & PA(deg)& $\sigma$ \\
\hline
4615& 0.00423 & 18.1 & 4.3 & $-$110.6 & 24.2 & $-$74.0 & 8.1 & ~50.8 & ~5.8 & $-$36.7& 4.3 
& $-$41.0 & 5.4   \\
4653& 0.00410 & 27.4 & 4.8 & $-$84.3 & 16.3 & $-$62.5 & 9.9 & ~72.1 & 25.6 & $-$10.4 & 5.0
& $-$25.6 & 6.2   \\
4850& 0.00384 & 37.4 & 4.7 & $-$67.9 & ~7.4 & $-$49.6 & 4.8 & ~73.8 & ~9.2 & ~7.9 & 4.2 & $-$13.3
& 4.1   \\
5090& 0.00348 & 44.7 & 4.8 & $-$63.1 & 10.0 & $-$42.2 & 4.7 & ~63.5 & ~9.8  & $-$5.8 & 4.8 & ~1.4
& 4.8  \\
8405& 0.00127 & 216.6 & 16.4 & ~$-$6.7 & ~3.1 & ~158.6 & 8.6 & 135.6 & ~2.2 & 129.6 & 0.7 & 170.8
& 7.2  \\
\hline
\end{tabular}
\end{table*}
The PA vs $\lambda^2$ plots for the selected regions are presented in
Fig.\,6 for the source 0548+165, and in Fig.\,7 for the source 
1524$-$136. 
In Table\,7, the derived RMs, the intrinsic PAs at the point of emission 
plus the RM in the rest frame ($rf$,
i.e. corrected for the  redshift) are presented for the two sources.
In each plot the best fit to the data is shown.
The RMs are estimated by fitting the points with a linear least-squares
fit. The correlation coefficients are $\geq 0.98$ in all cases but for regions
B and D in 1524$-$136 where we found 0.95. Subtracting $180^\circ$ from the
PAs of the 8.4\,GHz point for the six regions selected in 1524$-$136, 
the correlation 
coefficients drop to values $< 0.61$ in four cases. We found 0.82 for 
region B and 0.94 for region D. In conclusion, only in region D do
the RM values reported in Table\,7 present an ambiguity.
%
%
%
\begin{figure*}
\centering
\includegraphics[width=0.50\textwidth,bb=100 100 430 720,clip]{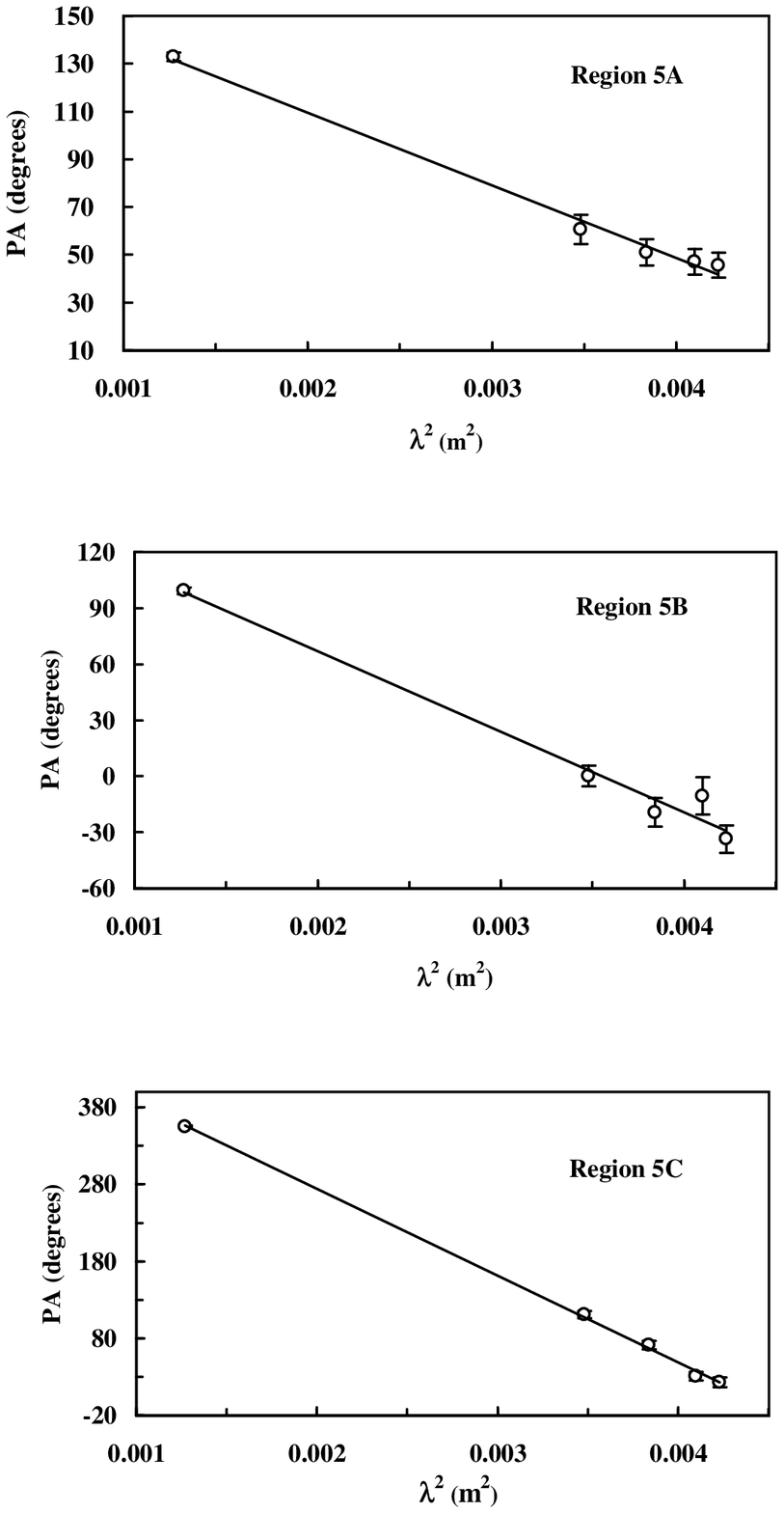}
\caption{0548+165: PA vs $\lambda^2$ plots. \label{ }} 
\end{figure*}
%
%
\begin{figure*}
\centering
\includegraphics[width=1.00\textwidth,bb=55 260 500 730,clip]{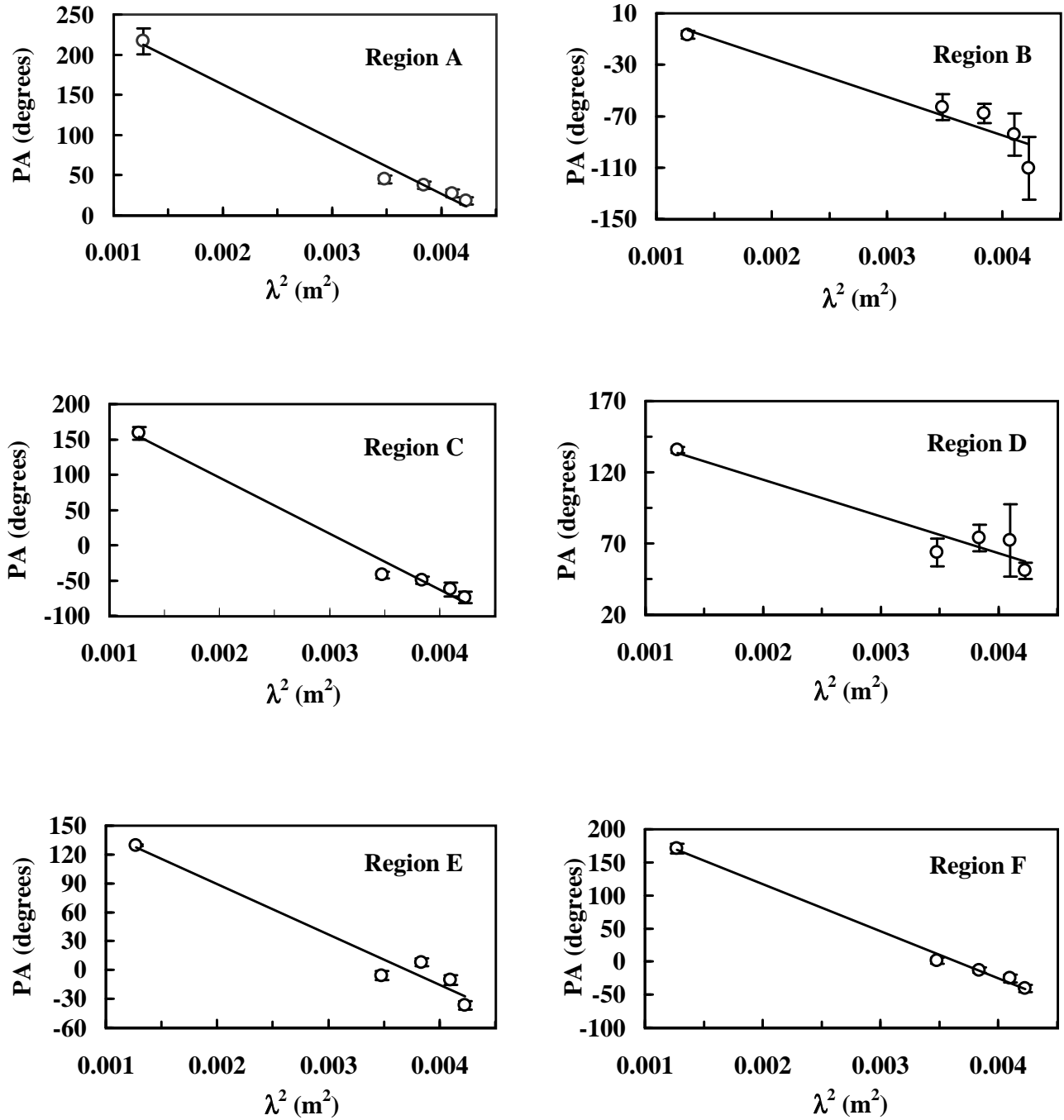}
\caption{1524$-$136: PA vs $\lambda^2$ plots.} 
\end{figure*}
%
%
\begin{table*}
\centering
\caption{\bf RM, PA$_{intrinsic}$ and RM$_{rf}$ for 0548+165 and
1524$-$136. \label{allrms} }
\vspace{0.5cm}          
\begin{tabular}{lcccc}
\hline
Source & Region&     RM         &PA$_{intrinsic}$     & RM$_{rf}$  \\
       &       &(rad m$^{-2}$)  &    (deg)        & (rad m$^{-2}$) \\
\hline
0548+165& 5A  &  $-$530$\pm$26  & 170.2$\pm$5.4   & $-$1150$\pm$56        \\
        & 5B  &  $-$753$\pm$65  & 153.1$\pm$13.3  & $-$1636$\pm$141   \\
        & 5C  &  $-$1968$\pm$36 & 140.0$\pm$7.4   & $-$4275$\pm$78    \\
\hline						                   
1524$-$136&A  &  $-$1185$\pm$85 & 299.0$\pm$17.4  & $-$8556$\pm$613    \\
          &B  &  $-$522$\pm$99  & 34.7$\pm$20.1   & $-$3769$\pm$715    \\
          &C  &  $-$1385$\pm$99 & 254.5$\pm$20.1  & $-$10000$\pm$715   \\
          &D  &  $-$448$\pm$82  & 165.9$\pm$16.7  & $-$3235$\pm$592    \\
          &E  &  $-$913$\pm$112 & 193.9$\pm$22.7  & $-$6592$\pm$809    \\
          &F  &  $-$1239$\pm$52 & 258.7$\pm$10.6  & $-$8946$\pm$375    \\

\hline
\end{tabular}
\end{table*}
\section{Discussion}
\subsection{0548$+$165}

The nucleus of 0548+165 is almost certainly component 1 (see Fig.\,2).
Using the flux density values derived from the 1.6-GHz observations
by Mantovani et al. (1998) and from the present observations at 5
and 8.4\,GHz (Table~3), a spectral index of $\alpha\leq 0.2$ is
found for component~1.  The remaining 
components have steeper spectral indices.  The projected linear size for
0548$+$165
from the nucleus to the region where the
jet shows an $\sim 90^\circ$ bend
is $\sim$330 pc. 
Regions, 5A, 5B and 5C all
show large RMs, much larger than the typical
RM$\leq$100 rad m$^{-2}$ due to the ISM of our Galaxy in this
direction (Simard-Normandin, Kronberg  \& Button 1981).

The  magnetic field direction derived from the observations lies
parallel to the tangent of the total intensity contours as we move
around the $\approx90^{\circ}$ bend.  This suggests that the field
lines are compressed by an interaction of the jet with a dense
cloud.  It is also of interest that this interaction does not
appear to significantly disrupt or perturb the jet from a smooth
flow.  This is also supported by sensitive (Mk3, Mode A)
$\lambda6$-cm VLBI observations (Mantovani et al. 1998).  

The largest value of RM is found in region 5C, where the jet is most
bent.  The value of RM then decreases going outwards along the
jet from regions 5B to 5A. In the three regions, the PAs are
rotated by angles larger than $\pi$/2 by 6-cm band. In addition, in
the region where the source structure suggests that an interaction
between the jet and the ISM is taking place, the
depolarization between 8.4 and 4.6\,GHz is p$_{4.6}$/p$_{8.4}$=0.26.
This indicates (Burn 1966) that the rotation is due to a `thin
slab' plus something else, such as a two-phase foreground medium,
i.e. narrow line clouds and a hot medium are present.

The polarized signal lies predominantly on the western edge of the
bent jet i.e. on the side where the putative collision occurs.  This
may be explained simply by compression of magnetic field lines by
the collision, with consequent enhancement of both the total
intensity and the linearly polarised emission.  

The depolarization produced by a Faraday screen is given by:
\begin{equation}
p(\lambda) = (1+8\sigma ^2 \lambda ^4 t^2 / s_{\circ}^{2})^{-1/2}
\end{equation}
(Tribble 1992) where $p$ is the fractional polarization, $\lambda$ the
wavelength, $\sigma$ the RM dispersion assuming a stochastic Gaussian
distribution, $t$ the resolution, and $s_\circ$ the RM scale fluctuation. 

Assuming $s_\circ$ to be of the order of the separation between the peaks of 
polarized emission ($\sim7.5$ mas), $\sigma=$ 1372 rad m$^{-2}$, 
i.e. the dispersion in the rest frame, and $t =$ 7 mas,
a value p$_{4.6}$/p$_{8.4}$ = 0.32 for the depolarization is obtained, in
good agreement with the value measured in the area of the bend 
(see Table\,3). It follows that the observed depolarization might be due 
to beam depolarization since the RM scale fluctuation is of the order of the
resolution achieved by the present observations.

The RM values in regions 5A and 5B are similar (at the 3-$\sigma$ level) 
and can be
understood as being due to an external screen. In region 5C, the
RM excess might be due to the presence of a NLR, which could also
 be responsible for the large bend in the jet. In fact, 3-D
simulations of jet-cloud interactions (Norman \& Balzara 1993) have
shown that a jet can be deflected by angles up to 90$^\circ$ preserving
its collimation after the bend.  De Young (1991) showed that a jet can
be deflected by interacting with a cloud in the NLR on a time scale
of $\sim10^6$ yr at a distance of 2--4 kpc from the nucleus.

Making the assumptions that the external screen is responsible for the RM
of  regions 5A and 5B, and partially in region 5C, and that the magnetic 
field, $B$, along
the line of sight and the thermal electron density, $n_e$, are almost constant,
the rotation measure is RM = $ 0.8 n_e B_\parallel l$. 
With RM = 1200 rad m$^{-2}$, and a  $l$ = 1\,kpc thick screen, the product
$n_e B_\parallel$ is 1.5 cm$^{-3} \mu$G. The typical range for the density
of thermal electron is $n_e \simeq$ 0.1 -- 1\,cm$^{-3}$, suggesting that
$B_\parallel \simeq 1.5 - 15 \mu$G.

Similar values for the magnetic field were derived by Venturi \& Taylor (1999)
for the bending jet of 3C216. 

\subsection {1524$-$136}
The region of jet-cloud interaction in the case of 1524$-$136 is situated
475\,pc from the nucleus. The RM value changes going from component A to E 
making the 
astrophysical analysis more intriguing. The peak of polarized emission
is found in region A. The jet changes direction in region B where, however,
the RM is lower. Concerning depolarization, this is seen for regions 
 A$+$B and C, while region F shows no depolarization.

Following Tribble (1992), 1524$-$136 has a depolarization 
p$_{4.6}$/p$_{8.4}$ = 0.33 assuming $s_\circ = 24$\,mas (mean distance between
components), $t =$ 6\,mas and $\sigma =$ 2577 rad m$^{-2}$. Integrating the 
total flux and the polarized flux over the whole source, a similar value for
the depolarization is found, i.e. p$_{4.6}$/p$_{8.4}$ = 0.28. 
This result might
suggest that the depolarization can be due to inhomogenities in the 
structure of the
magnetic field responsible for the changes in the RM, caused by the external
screen, on a comparable scale to that of the distances between the components.
Since the differences between PAs is $> 90^\circ$, we can say that the RM
is due to an external screen in which the orientation of the magnetic field
along the line of sight and/or the electron density, do vary along
the jet in such a way that the observed variation in RMs can be reproduced.

The intensity of the magnetic field along the line of sight for each region,
assuming $n_e\simeq 0.1 - 1$ cm$^{-3}$ and a 1 kpc thick screen, is
summarized in Table\,8. The values obtained for $B_\parallel$ in 1524$-$136 
are larger than those found for the screen in 0548$+$165.

\begin{table}
\begin{center}
\caption{Line-of-sight magnetic field intensity B$_{\parallel}$ in regions
of 0548$+$165 and 1524$-$136}
\centering
\begin{tabular}{lcc}
\hline
Source     & Region    &  B$_\parallel$ ($\mu$G)  \\   
\hline
0548$+$165 & 5(A+B+C)     & 15 -- 2                      \\
1524$-$136 &  A        & 107 -- 11                        \\
           &  B        & 47 -- 5                        \\
           &  C        & 125 -- 13                        \\
           &  D        & 40 -- 4                        \\
           &  E        & 82 -- 8                        \\
           &  F        & 112 -- 11                        \\

\hline
\end{tabular}
\end{center}
\end{table}
\section{Conclusions}

In this paper we have presented results obtained for the CSSs
0548+165 and 1524$-$136 from VLBA+VLA1 observations at five
frequencies.  These sources were the first observed in a larger
project planned to study CSSs which have polarization
percentages that decrease with decreasing frequency and high
values of rotation measure (RM$>$450 rad m$^{-2}$) on the arcsecond
scale.

Both sources show polarized emission along their mas-scale jets, while 
weak polarized emission is detected in their cores. We were able to estimate
rotation measures by a five-point linear fit to
PA vs $\lambda^2$ plots in three and six regions along the jets of
0548+165 and 1524$-$136 respectively. The RMs obtained show very high
values in both sources (RM$>$ 4000 and up to 10$^4$
rad m$^{-2}$ respectively). We suggest that the high RMs are
produced by a foreground screen in 1524$-$136 and by a foreground
screen plus a RM excess due to the effect of a NLR cloud for 0548+165.
In recent HST optical images of CSSs, a central compact component of
$<$300\,mas, dominated by forbidden
[OIII]$\lambda\lambda4959,5007$\,\AA  ~and [OIII]$\lambda3727$\,\AA
~emission lines, is detected. Compact emission line regions with
dimensions typical of the NLR of Seyfert galaxies are also observed
(Axon et al. 2000).

Estimates of the magnetic field intensity along the line 
of sight in the foreground screen responsible for Faraday rotation have been
made on the basis of thermal electron density and screen depth. 
Assuming $n_e \simeq 0.1\ \mbox{cm}^{-3}$ for the thermal electron density 
and $l \simeq 1\ \mbox{kpc}$ for the Faraday screen depth, we find a magnetic
field intensity value of $\sim$ 15 $\mu$G in the foreground screen. 
A magnetic field in the range 40--125 $\mu$G was found in 1524$-$136.
The electron density and Faraday-screen depth estimates used, are uncertain 
(by a factor of ten) because of lack of information on X--ray emission and 
on the position angle between the jet axis and the observer's line of sight. 

The observed depolarization is consistent with that derived using the 
Tribble (1992) model which analyzes the effects of a Faraday
screen. This suggests that the depolarization is probably due  
to beam depolarization in 0548+165 and most likely due to inhomogeneities 
in the magnetic field structure of the foreground screen in 1524$-$136.

The derived magnetic field directions follow the jets in both sources, 
even when
the source is strongly bent due to the interaction with the dense external
medium. However, the jets maintain their collimation despite a possible
strong interaction with a NLR.

Finally, we note the difference between the distribution of polarised
emission in this pair of two jet-dominated QSOs and that found in 
core-dominated 
QSOs. For the latter class, the highest RMs are found
in the inner 20 pc of the nuclear region (Taylor 1999). In the
CSSs 0548+165 and 1524$-$136 the most polarised regions are along the jet 
while the nuclei are weakly polarised.
\begin{acknowledgements}
The authors whish to thank the referee Dr. G.B. Taylor for valuable
comments on the manuscript.
\end{acknowledgements}

\end{document}